\documentclass[
aps,%
12pt,%
final,%
notitlepage,%
oneside,%
onecolumn,%
nobibnotes,%
nofootinbib,%
superscriptaddress,%
showpacs,%
centertags]%
{revtex4}

\newtheorem{defin}{Definition}
\newtheorem{theor}{Theorem}
\newtheorem{prop}{Proposition}
\newtheorem{cor}{Corollary}
\newenvironment{demo}
{\bgroup\par\smallskip\noindent{\it Proof: }}{\rule{0.5em}{0.5em}
\egroup}
\def\bb {\begin {eqnarray}}
\def\ee {\end {eqnarray}}

\begin{document}\selectlanguage{english}
\title
{Hidden symmetries and Killing tensors \\
on curved spaces 
}
\author{\firstname{Stere}~\surname{Ianu\c s }}
\affiliation{Department of Mathematics, University of Bucharest,
         Bucharest, Romania }
\author{\firstname{Mihai}~\surname{Visinescu}}
\affiliation{Department of Theoretical Physics,\\
Institute for Physics and Nuclear Engineering,
Magurele, Bucharest, Romania }
\author{\firstname{Gabriel-Eduard}~\surname{V\^\i lcu }}
\affiliation{Department of Mathematics and Computer Science,\\ \
Petroleum-Gas University of Ploie\c sti,
Ploie\c sti, Romania }

\begin{abstract}
Higher order symmetries corresponding to Killing tensors are 
investigated.
The intimate relation between Killing-Yano tensors and non-standard 
supersymmetries is pointed out. In the 
Dirac theory on curved spaces, Killing-Yano tensors generate Dirac 
type operators involved in interesting algebraic structures as 
dynamical algebras or even infinite dimensional algebras or superalgebras. 
The general results are applied to space-times which appear in modern 
studies. 
One presents the infinite dimensional superalgebra of Dirac type 
operators on the 4-dimensional Euclidean Taub-NUT space that can be 
seen as a twisted loop algebra. 
The existence of the conformal Killing-Yano tensors is investigated for 
some spaces with mixed Sasakian structures.
\end{abstract}

\maketitle

\section{Introduction}\label{int1}

One of the key concepts in physics is that of symmetries, Noether's 
theorem giving a correspondence between symmetries and conserved 
quantities. For the geodesic motions on a space-time the usual 
conserved quantities are related to the isometries which
correspond to Killing vectors. Sometimes a space-time could admit 
higher order symmetries described by symmetric Killing tensors, called 
St\" ackel-Killing (S-K).
These symmetries are known as {\it hidden 
symmetries} and the typical example is the  Runge-Lenz vector in 
the Kepler/Coulomb problem. The corresponding conserved quantities are 
quadratic, or, more general, polynomial in momenta.
Their existence guarantees 
the integrability of the geodesic motions and is intimately related to 
separability of Hamilton-Jacobi (see, e. g. \cite{BEN}) and the 
Klein-Gordon equation at the quantum level \cite{Carter}.

The next most simple objects that can be studied in connection with 
the symmetries of a manifold after the S-K tensors are the 
Killing-Yano tensors (K-Y) \cite{YAN}. 
It was observed \cite{GRH} that a K-Y tensor 
generates additional {\it supercharges} in the dynamics of 
pseudo-classical spinning particles being the
natural geometrical objects to be coupled with the fermionic degrees 
of freedom \cite{GRH,Car}. 
In this way it was realized the significant
connection between K-Y tensors and {\it non-standard 
supersymmetries}. Passing to quantum Dirac equation it was discovered 
\cite{CML} that K-Y tensors generate conserved {\it 
non-standard} Dirac operators which commute with the {\it standard} 
one.

The conformal extension of the Killing tensor equation determines the 
conformal Killing tensors \cite{JJ} which define first integrals of the null 
geodesic equation. Investigation of the properties of higher-dimensional 
space-times has pointed out the role of the conformal K-Y 
(CKY)
tensors to generate background metrics with black-hole solutions (see, 
e. g. \cite{FRO}).

The aim of this paper is to investigate a few examples of 
curved spaces endowed with special structures admitting
K-Y tensors
which could be relevant in the theories of modern physics.

The first example is represented by the 4-dimensional Euclidean 
Taub-Newman-Unti-Tamburino (Taub-NUT) space. The
motivation to carry out this example is twofold. First of all, in the 
Taub-NUT geometry there are known to exist four K-Y tensors 
\cite{GR}. From this point of view the  Taub-NUT manifold is an 
exceedingly interesting space to exemplify the effective construction of 
the conserved quantities in terms of geometric ones. On the other hand, 
the Taub-NUT geometry is involved in many modern studies in physics. 

In the second example we investigate the existence of CKY tensors in 
higher dimensional space-times \cite{IVV}.
Investigations of the properties of space-times of higher dimensions 
($D > 4$) have recently attracted considerable attention as a result of 
their appearance in theories of unification such as string and $M$ 
theories.
Versions of $M-$theory could be formulated in space-times with various 
number of time dimensions giving rise to exotic space-time signatures. 
The $M-$theory in $10+1$ dimensions is linked via dualities to a $M^{*}$ 
theory in $9+2$ dimensions and a $M^{'}$ theory in $6+5$ dimensions. 
Various  limits of these will give rise to $IIA-$ and $IIB-$like string 
theories in many variants of dimensions and signatures \cite{HUL}.

The plan of the paper is as follows: In section \ref{sec2} we present the 
Killing tensors which generalize the Killing vectors.
In section \ref{sec3} we describe the Dirac-type operators generated by K-Y 
tensors and the general results are applied to the 
4-dimensional Euclidean Taub-NUT space \cite{Taub,NUT}. 
The CKY tensors on manifolds with 
mixed $3-$structures are presented in the last section \ref{sec4}.
Some details concerning the geometrical properties of 
these metrics are summarized in the Appendices.

\section{Killing vector fields and their generalizations}\label{sec2}

Let $(M,g)$ be a semi-Riemannian manifolds. A vector field $X$ on
$M$ is said to be a Killing vector field if the Lie derivative with
respect to $X$ of the metric $g$ vanishes.

Killing vector fields can be
generalized to conformal Killing vector fields \cite{YAN}, i.e.
vector fields with a flow preserving a given conformal class of
metrics.
A natural generalization of conformal Killing vector fields is given
by the CKY tensors \cite{KSW}.

\begin{defin}
A CKY tensor of rank $p$ on a semi-Riemannian
manifold $(M,g)$ is a $p$-form $f$ which satisfies:
\begin{equation}\label{4}
\nabla_X f=\frac{1}{p+1}X \lrcorner
f -\frac{1}{n-p+1}X^*\wedge d^*f \,,
\end{equation}
for any vector field $X$ on $M$.
\end{defin}

Here $\nabla$ is the Levi-Civita
connection of $g$, $n$ is the dimension of $M$, $X^*$ is the 1-form
dual to the vector field $X$ with respect to the metric $g$,
$\lrcorner$ is the operator dual to the wedge product and $d^*$ is
the adjoint of the exterior derivative $d$. If $f$ is co-closed
in (\ref{4}), then we obtain the definition of a K-Y tensor
(introduced by Yano \cite{YAN}). We can easily see that for $p=1$,
they are dual to Killing vector fields.

A K-Y tensor can be characterized in several ways. In a equivalent 
manner a differential $p-$form $f$ is called a K-Y tensor if its 
covariant derivative $\nabla_\lambda f_{\mu_1...\mu_p}$ is totally 
antisymmetric. As a consequence of the antisymmetry  a K-Y tensor
satisfy the equation
$\nabla_{(\lambda} f_{\mu_1)...\mu_p} = 0$.
Let us remark that for covariantly constant K-Y tensors each term of 
the l. h. s. of this equation vanishes. The covariantly constant K-Y 
tensors
represent a particular class of K-Y tensors and they play a special
role in the theory of Dirac operators as it will be seen in 
section \ref{sec3}.

We mention that K-Y tensors are also called Yano tensors or
Killing forms, and CKY tensors are sometimes
referred as conformal Yano tensors, conformal Killing forms or
twistor forms \cite{BMS, MS, SEM}.

For generalizations of the Killing vectors one might also consider 
higher order symmetric tensors.

\begin{defin}
A symmetric tensor  of $ K_{\mu_1...\mu_r} $ of rank $r>1$ 
satisfying a generalized Killing equation
$\nabla_{(\lambda} K_{\mu_1...\mu_r)} = 0$
is called a S-K tensor.
\end{defin}

The relevance in physics of the S-K tensors is given by the following 
proposition which could be easily proved:

\begin{prop}
A symmetric tensor $K$ on $M$ is a S-K tensor iff the 
quantity
$K = K_{\mu_1...\mu_r} \dot{s}^{\mu_1}\cdots \dot{s}^{\mu_r}$
is constant along every geodesic $s$ in $M$.
\end{prop}

Here the over-dot denotes the ordinary proper time derivative and the 
proposition ensures that $K$ is a first integral of the geodesic 
equation.

These two generalizations S-K and  K-Y of the Killing 
vectors could be related. Let $f_{\mu_1...\mu_p}$ be a K-Y 
tensor, 
then the tensor field
$K_{\mu\nu} =  f_{\mu\mu_2...\mu_p} f^{\mu_2...\mu_p}_{~~~~~~~\nu}$
is a S-K tensor and it sometimes refers to this S-K tensor as the 
associated tensor with $f$.
However, the converse statement is not true in general: not all S-K 
tensors of rank $2$ are associated with a K-Y tensor. 

\section{Dirac-type operators}\label{sec3}

For a quantum relativistic description of a spin-1/2 particle on a 
curved space we use the {\it standard} Dirac operator 
\begin{equation}\label{DS}
D_s = i  \gamma^\mu \nabla_\mu \,,
\end{equation}
where $\nabla_\mu$ are the spin covariant derivatives including 
spin-connection, while $\gamma^\mu$ are the standard Dirac matrices 
carrying natural indices.

We note that for any isometry with a Killing vector $R^\mu$ 
there is an appropriate operator 
\begin{equation}\label{vk}
X_k = - i ( R^\mu \nabla_\mu - \frac{1}{4} \gamma^\mu \gamma^\nu 
R_{\mu;\nu}) \,,
\end{equation}
which commutes with $D_s$.
Moreover each K-Y tensor 
$f_{\mu\nu}$ produces a {\it non-standard} Dirac operator of the form
\cite{CML}
\begin{equation}\label{df}
D_f = i\gamma^\mu (f_\mu ^{~\nu}\nabla_\nu  - 
\frac{1}{6}\gamma^\nu 
\gamma^\rho f_{\mu\nu;\rho}) \,,
\end{equation}
which anticommutes with the standard Dirac operator $D_s$
and can be involved in new types of genuine or hidden 
(super)symmetries.

\subsection{Covariantly constant K-Y tensors}

Remarkable superalgebras of Dirac-type operators can be produced by 
special 
second-order K-Y tensors that represent square roots of the metric 
tensor \cite{Kl,NOVA,CV2}.

\begin{defin}\label{Def1}
The  non-singular real or complex-valued K-Y tensor $f$ of rank 2 
defined on
$M$ which satisfies
\begin{equation}\label{ffg}
f^{\mu}_{~\alpha} f_{\mu\beta}=g_{\alpha\beta}\,,
\end{equation}
is called an unit root of the metric tensor of $M$, or simply an unit 
root of $M$.
\end{defin}

Let us observe that any unit root K-Y tensor  is 
covariantly constant \cite{Kl}, i. e. 
$f_{\mu\nu;\sigma}=0\,.$

It is worthy to be noted that the covariantly constant K-Y tensors give 
rise  to  Dirac-type operators 
of the form (\ref{df}) connected with the standard Dirac operators as 
follows:

\begin{theor}\label{DtyD}
The Dirac-type operator $D_f$ produced by the K-Y tensor $f$
satisfies the  condition
$(D_{f})^2=D^2_s$
iff $f$ is an unit root.
\end{theor}
\begin{demo}
The arguments of Ref.  \cite{Kl} show that the condition   
from the theorem is
equivalent  with   (\ref{ffg}) $f$ being a covariantly constant 
K-Y tensor.
\end{demo}

\subsection{Dirac operators on Euclidean Taub-NUT space}

To make things more specific let us consider the 
Euclidean Taub-NUT space (see Appendix A) which 
is hyper-K\" ahler and possesses many non-standard symmetries expressed 
in terms of four K-Y tensors and three S-K tensors.

From the covariantly constant K-Y tensors $f^i$ (\ref{fi}),
using prescription (\ref{df}),
we can construct three Dirac-type operators $D^{(i)}$ which 
anticommute with standard Dirac operator $D_s$ (\ref{DS}). It is 
convenient to define \cite{cv08}
$Q_i = i H^{-1} D^{(i)}$
where  $H = - \gamma^0 D_s$ is the massless Hamiltonian operator. 
These operators form a representation of the quaternionic units:
$Q_{i}\, Q_{j}=\delta_{ij}I+i\varepsilon_{ijk} Q_{k}$.

On the other hand Dirac-type operator constructed from the K-Y tensor 
$f^Y$ (\ref{fy}) is $D^Y$ and again it is convenient to define a new 
operator $Q^Y = H D^Y$.

The conserved Runge-Lenz operator of the Dirac theory is 
\begin{equation}\label{rl}
{\cal K}_{i}=\frac{\mu}{4}\{ Q^Y,\, Q_i\} + \frac{1}{2}({\cal 
B}-P_4)Q_i-{\cal
J}_i  P_4 \,,
\end{equation}
where ${\cal B}^{2}={P_{4}}^{2}-H^2$, $J_i \,,(i=1,2,3)$ are the 
components of the total angular momentum, while $P_4 = - i \partial_4$
corresponding to the fourth Cartesian coordinate 
$x^4 = - 4 m (\chi + \varphi)$.

The operators $J_i$ and $ {\cal K}_{i}$ are involved in the following 
system of commutation relations:
\begin{equation}\label{jjkk}
\left[ {\cal J}_{i},\; {\cal J}_{j}\right]=i\varepsilon
_{ijk}{\cal J}_{k}
,\;
\left[ {\cal J}_{i},\, {\cal K}_{j}\right]=i\varepsilon
_{ijk}{\cal K}_{k}
\,,\qquad
\left[ {\cal K}_{i},\, {\cal K}_{j}\right]=i\varepsilon
_{ijk}{\cal J}_{k} {\cal B}^{2}\,,
\end{equation}
and commute with the operators $Q_i$ 
\begin{equation}\label{jkq}
\left[ {\cal J}_{i},\, {Q}_{j}\right]=i\varepsilon _{ijk}{Q}_{k}
\,,\qquad \left[ {\cal K}_{i},\, {Q}_{j}\right]=i\varepsilon
_{ijk}{Q}_{k} {\cal B}\,.
\end{equation}
The algebra (\ref{jjkk}) does not close as a finite Lie algebra because
of the factor ${\cal B}^2$. 
In the standard treatment one concentrates on individual subspaces of 
the whole Hilbert space which belong to definite eigenvalues of 
${\cal B}^{2}$. This is similar to the dynamical algebra of the 
hydrogen atom  which can be identified in a natural way 
with an infinite dimensional twisted loop algebra \cite{Daboul}.

The dynamical algebras of the Dirac theory 
have to be obtained by replacing this operator ${\cal B}^2$ with its 
eigenvalue $q^2 - E^2$ and rescaling the operators ${\cal K}_{i}$. The 
same kind of problems appears for the anticommutators involving the 
fermionic operators $Q_i$ and $Q^Y$. In what follows, in order to keep 
the presentation as simple as possible, we shall only give the briefest 
account of the algebra of operators connected with hidden symmetries in 
the bosonic sector. For the algebra of operators from the fermionic 
sector the reader should consult \cite{CV1}.

In the bosonic sector of conserved operators let us define the new 
operators "absorbing" the operator ${\cal B}$ by assigning 
grades to each operator \cite{mv08}:
\begin{equation}
A_{2n}^i:={\cal J}_i{\cal B}^n\,,\quad 
B_{2n+2}^i:={\cal K}_i{\cal B}^n\,,
\end{equation}
for any $n= 0,1,2...$. 
The algebra of these operators can be seen as 
an infinite dimensional twisted loop 
algebra of the Kac-Moody type. 
In this way the commutation relations of the bosonic sector are
given by a Kac-Moody type algebra:
\begin{eqnarray}\label{KM}
\left[ {A}^{i}_{2n}, {A}^{j}_{2m}\right]
&=&i\varepsilon_{ijk}{A}^{k}_{2(n+m)}\,, \nonumber\\
\left[ {A}^{i}_{2n}, {B}^{j}_{2m+2}\right]
&=&i\varepsilon_{ijk}{B}^{k}_{2(n+m+1)}\,,\\
\left[ {B}^{i}_{2n+2}, {B}^{j}_{2m+2}\right]
&=&i\varepsilon_{ijk}{A}^{k}_{2(n+m+2)}\, . \nonumber
\end{eqnarray}

\section{CKY tensors on manifolds with mixed 3-structures}\label{sec4}

An almost para-hypercomplex structure on a smooth manifold $M$ is a
triple $H=(J_{\alpha})_{\alpha=\overline{1,3}}$, where $J_1$ is an
almost complex structure on $M$ and $J_2$, $J_3$ are almost product
structures on $M$, satisfying: $J_1J_2J_3=-Id$. In this case $(M,H)$
is said to be an almost para-hypercomplex manifold.

A semi-Riemannian metric $g$ on $(M,H)$ is said to be
para-hyperhermitian if it satisfies
$g(J_\alpha X,J_\alpha Y)=\epsilon_{\alpha}
g(X,Y),\ \alpha\in\{1,2,3\}$
for all $X,Y\in\Gamma(TM)$, where $\epsilon_1=1,
\epsilon_2=\epsilon_3=-1$. In this case, $(M,g,H)$ is called an
almost para-hyperhermitian manifold. Moreover, if each $J_\alpha$ is
parallel with respect to the Levi-Civita connection of $g$, then
$(M,g,H)$ is said to be a para-hyper-K\"{a}hler manifold.

\begin{theor} Let $(M,g)$ be a semi-Riemannian manifold. Then the following
five assertions are mutually equivalent:

(1) $(M,g)$ admits a mixed 3-Sasakian structure.

(2) The cone $(C(M),\overline{g})=(M\times\mathbf{
R}_+,dr^2+r^2g)$
admits a para-hyper-K\"{a}hler structure.

(3) There exists three orthogonal Killing vector fields
$\{\xi_1,\xi_2,\xi_3\}$ on $M$, with $\xi_1$ unit space-like vector
field and $\xi_2, \xi_3$ unit time-like vector fields satisfying
\begin{equation}\label{17}
[\xi_\alpha,\xi_\beta]=-2\epsilon_\gamma\xi_\gamma,
\end{equation}
where $(\alpha,\beta,\gamma)$ is an even permutation of (1,2,3) and
$\epsilon_1=1, \epsilon_2=\epsilon_3=-1$, such that the tensor
fields $\phi_\alpha$ of type (1,1), defined by: $\phi_\alpha
X=-\epsilon_{\alpha}\nabla_X\xi_\alpha$, $\alpha\in\{1,2,3\}$, 
satisfies the conditions (\ref{12}), (\ref{13}) and (\ref{14}).

(4) There exists three orthogonal Killing vector fields
$\{\xi_1,\xi_2,\xi_3\}$ on $M$, with $\xi_1$ unit space-like vector
field and $\xi_2, \xi_3$ unit time-like vector fields satisfying
(\ref{17}), such that:
\begin{equation}
R(X,\xi_\alpha)Y=g(\xi_\alpha,Y)X-g(X,Y)\xi_\alpha,\
\alpha\in\{1,2,3\},
\end{equation}
where $R$ is the Riemannian curvature tensor of the Levi-Civita
connection $\nabla$ of $g$.

(5) There exists three orthogonal Killing vector fields
$\{\xi_1,\xi_2,\xi_3\}$ on $M$, with $\xi_1$ unit space-like vector
field and $\xi_2, \xi_3$ unit time-like vector fields satisfying
(\ref{17}), such that the sectional curvature of every section
containing $\xi_1, \xi_2$ or $\xi_3$ equals 1.
\end{theor}

\begin{demo}
$(1)\Rightarrow(2)$ If $M^{4n+3}$ is  a manifold
endowed with a mixed 3-Sasakian structure (see Appendix B)
$((\phi_\alpha,\xi_\alpha,\eta_\alpha)_{\alpha=\overline{1,3}},g)$,
then we can define a para-hyper-K\"{a}hler structure
$\{J_\alpha\}_{\alpha=\overline{1,3}}$ on the cone
$(C(M),\overline{g})=(M\times\mathbf{R}_+,dr^2+r^2g)$, by
$J_\alpha X=\phi_\alpha X-\eta_\alpha(X)\Phi, J_\alpha\Phi=\xi_\alpha$
for any $X\in\Gamma(TM)$ and $\alpha\in\{1,2,3\}$, where
$\Phi=r\partial_r$ is the Euler field on $C(M)$.

$(2)\Rightarrow(1)$ If the cone
$(C(M),\overline{g})=(M\times\mathbf{R}_+,dr^2+r^2g)$ admits a
para-hyper-K\"{a}hler structure
$\{J_\alpha\}_{\alpha=\overline{1,3}}$, then we can identify $M$
with $M\times\{1\}$ and we have a mixed 3-Sasakian structure
$((\phi_\alpha,\xi_\alpha,\eta_\alpha)_{\alpha=\overline{1,3}},g)$
on $M$ given by
$\xi_\alpha=J_\alpha(\partial_r),\, \phi_\alpha X=-\epsilon_{\alpha}
\nabla_X \xi_\alpha,\, \eta_\alpha(X)=g(\xi_\alpha,X),$
for any $X\in\Gamma(TM)$ and $\alpha\in\{1,2,3\}$.

$(2)\Leftrightarrow(3)$ This equivalence is clear (see also
\cite{BG}).

$(3)\Leftrightarrow(4)$ This equivalence follows from direct
computations.

$(4)\Leftrightarrow(5)$ This equivalence follows using the formula
of the sectional curvature.
\end{demo}

From the above Theorem we can easily obtain the next properties (see
also \cite{SEM}).

\begin{cor} Let $M^{4n+3}$ be a manifold endowed with a mixed 3-Sasakian
structure
$((\phi_\alpha,\xi_\alpha,\eta_\alpha)_{\alpha=\overline{1,3}},g)$.
Then:

(1) $\xi_1$ is unit space-like Killing vector field and
$\xi_2,\xi_3$ are unit time-like Killing vector fields on $M$;

(2) $\eta_1,\eta_2,\eta_3$ are CKY tensors of
rank 1 on $M$;

(3) $d\eta_1,d\eta_2,d\eta_3$ are CKY tensors of
rank 2 on $M$;

(4) $M$ admits K-Y tensors of rank $(2k+1)$, for
$k\in\{0,1,...,2n+1\}$.
\end{cor}
\begin{cor} Let $M^{4n+3}$ be a manifold endowed with a
mixed 3-Sasakian structure
$((\phi_\alpha,\xi_\alpha,\eta_\alpha)_{\alpha=\overline{1,3}},g)$.
Then the distribution spanned by $\{\xi_1,\xi_2,\xi_3\}$ is
integrable and defines a 3-dimensional Riemannian foliation on $M$,
having totally geodesic leaves of constant curvature 1.
\end{cor}

M. V. would like to thank the organizers for the excellent XXVII 
International Colloquium
{\it Group Theoretical Methods in Physics} held in Yerevan, 
Armenia in August 2008. 
The work of M. V. was supported in part by CNCSIS Programs, Romania.

\appendix
\setcounter{equation}{0} \renewcommand{\theequation}
{A.\arabic{equation}}
\section*{Appendix A. Euclidean Taub-NUT space}	

Let us consider the Taub-NUT space \cite{Taub,NUT} and the chart with Cartesian 
coordinates $x^\mu (\mu, \nu =1,2,3,4)$ having the line element
\begin{eqnarray}\label{metric}
ds^{2}= g_{\mu\nu}dx^{\mu}dx^{\nu}=f(r)(d\vec{x})^{2}
 + \frac{g(r)}{16 m^2}(dx^{4}+ A_{i}dx^{i})^{2} \,,
\end{eqnarray}
where $\vec{x}$ denotes the three-vector 
$
\vec{x} = (r,\theta,\varphi)\,,
(d\vec{x})^{2}=(dx^{1})^{2}+(dx^{2})^{2}+(dx^{3})^{2}
$
and $\vec{A}$ is the gauge field of a monopole
$
{\rm div}\vec{A}=0\,, \quad \vec{B}\,={\rm rot}\, 
\vec{A}=4m\frac{\vec{x}}{r^3}.
$
The real number $m$ is a parameter of the theory which enter in the 
form of the functions
$ f(r) = g^{-1}(r) =  \frac{4 m + r}{r} $
and the so called NUT singularity is absent if $x^4$ is periodic with 
period $16 \pi m$. Sometimes it is convenient to make the coordinate 
transformation 
$ x^4 = - 4 m (\chi + \varphi),  $
with $0\leq \chi < 4\pi$.

In the Taub-NUT geometry there are four Killing vectors \cite{GR}. 
Three Killing vectors
correspond to the invariance of the metric (\ref{metric}) under spatial
rotations, obeying an $SU(2)$ algebra, while the fourth
generates the $U(1)$ of $\chi$ translations, commuting
with the other Killing vectors. 

On the other hand in the Taub-NUT geometry there are known to exist 
four K-Y tensors of valence 2. The first three 
\begin{equation}
f^i =8m(d\chi + \cos\theta d\varphi)\wedge dx_i 
- \epsilon_{ijk}(1+\frac{4m}{r}) dx_j \wedge dx_k\, ,
i,j,k=1,2,3 \,, \label{fi}
\end{equation}
are covariantly constant, i.e. $\nabla_\mu f^{i}_{\nu\lambda} =0$.
The $f^i$ 
define three anticommuting complex structures of the Taub-NUT manifold, 
their components realizing the quaternion algebra
$f^i  f^j + f^j f^i = - 2 \delta_{ij},\; f^i  f^j - f^j f^i = - 2
\varepsilon_{ijk} f^k$.
The existence of these K-Y tensors is linked to the 
hyper-K\"ahler geometry of the manifold and shows directly the relation 
between the geometry and the $N = 4$ supersymmetric extension of the 
theory \cite{GRH}.

The fourth K-Y tensor is
\begin{eqnarray}\label{fy}
f^Y = 8m(d\chi + \cos\theta  d\varphi)\wedge dr 
+4r(r+2m)(1+\frac{r}{4m})\sin\theta  d\theta \wedge d\varphi \,,
\end{eqnarray}
having  a non-vanishing  covariant derivative
$
f^{Y}_{r\theta;\varphi} = 2(1+\frac{r}{4m})r\sin\theta.
$

\setcounter{equation}{0} \renewcommand{\theequation}
{B.\arabic{equation}}
\section*{Appendix B. Manifolds with mixed 3-structures}
Let $M$ be a differentiable manifold equipped with a triple
$(\phi,\xi,\eta)$, where $\phi$ is a a field  of endomorphisms of
the tangent spaces, $\xi$ is a vector field and $\eta$ is a 1-form
on $M$ such that:
$\phi^2=-\epsilon I+\eta\otimes\xi,\  \eta(\xi)=\epsilon.$
If $\epsilon=1$ then $(\phi,\xi,\eta)$ is said to be an almost
contact structure on $M$ (see \cite{BLR}), and if $\epsilon=-1$ then
$(\phi,\xi,\eta)$ is said to be an almost paracontact structure on
$M$ (see \cite{MTS}).

\begin{defin} \cite{IMV}
Let $M$ be a differentiable manifold which admits an almost contact structure
$(\phi_1,\xi_1,\eta_1)$ and two almost paracontact
structures $(\phi_2,\xi_2,\eta_2)$ and
$(\phi_3,\xi_3,\eta_3)$, satisfying the following
conditions:
$$
\eta_\alpha(\xi_\beta)=0\,,  \forall \alpha\neq\beta \,,\qquad
\phi_\alpha(\xi_\beta)=-\phi_\beta(\xi_\alpha)=\epsilon_\gamma\xi_\gamma,
$$
$$
\eta_\alpha\circ\phi_\beta=-\eta_\beta\circ\phi_\alpha=
\epsilon_\gamma\eta_\gamma \,,\qquad
\phi_\alpha\phi_\beta-\eta_\beta\otimes\xi_\alpha=
-\phi_\beta\phi_\alpha+\eta_\alpha\otimes\xi_\beta=\epsilon_\gamma\phi_\gamma,
$$
where $(\alpha,\beta,\gamma)$ is an even
permutation of (1,2,3) and $\epsilon_1=1, \epsilon_2=\epsilon_3=-1$.

Then the manifold $M$ is said to have a mixed 3-structure
$(\phi_\alpha,\xi_\alpha,\eta_\alpha)_{\alpha=\overline{1,3}}$.
\end{defin}

\begin{defin} If a manifold $M$ with a mixed
3-structure
$(\phi_\alpha,\xi_\alpha,\eta_\alpha)_{\alpha=\overline{1,3}}$
 admits a semi-Riemannian metric $g$ such that:
$$
g(\phi_\alpha X, \phi_\alpha Y)=\epsilon_\alpha
g(X,Y)-\eta_\alpha(X)\eta_\alpha(Y),\, g(X,\xi_\alpha)=\eta_\alpha(X) \,,
$$
for all $X,Y\in\Gamma(TM)$ and $\alpha=1,2,3$, then we say that $M$
has a metric mixed 3-structure and $g$ is called a compatible
metric. Moreover, if $(\phi_1,\xi_1,\eta_1,g)$ is a Sasakian
structure, i.e. (see \cite{BLR}):
\begin{equation}\label{12}
(\nabla_X\phi_1) Y=g(X,Y)\xi_1-\eta_1(Y)X \,,
\end{equation}
and $(\phi_2,\xi_2,\eta_2,g)$, $(\phi_3,\xi_3,\eta_3,g)$ are
LP-Sasakian structures, i.e. (see \cite{MTS}):
\begin{equation}\label{13}
(\nabla_X\phi_2) Y=g(\phi_2X,\phi_2Y)\xi_2+\eta_2(Y)\phi_2^2X,\,
\end{equation}
\begin{equation}\label{14}
(\nabla_X\phi_3) Y=g(\phi_3X,\phi_3Y)\xi_3+\eta_3(Y)\phi_3^2X,
\end{equation}
then
$((\phi_\alpha,\xi_\alpha,\eta_\alpha)_{\alpha=\overline{1,3}},g)$
is said to be a mixed Sasakian 3-structure on $M$.
\end{defin}

It is easy to see that any manifold $M$ with a mixed 3-structure
admits a compatible semi-Riemannian metric $g$. Moreover, the
signature of $g$ is $(2n+1,2n+2)$ and the dimension of the manifold
$M$ is $4n+3$. The main property of a manifold endowed with a mixed
3-Sasakian structure is the following (see \cite{IV}):

\begin{theor} Any $(4n+3)-$dimensional manifold endowed with a
mixed 3-Sasakian structure is an Einstein space with Einstein
constant $\lambda=4n+2$.
\end{theor}

Concerning the symmetric Killing tensors let us note that
D.E. Blair studied in \cite{BL} the almost contact manifold with
Killing structure tensors. He assumed that $M$ has an almost contact
metric structure $(\phi,\xi,\eta,g)$ such that $\phi$ and $\eta$ are
Killing. Then he proved that if $(\phi,\xi,\eta,g)$ is normal, it is
a cosymplectic structure.

For a mixed 3-structure with a compatible semi-Riemannian metric
$g$, we have the following result:
\begin{prop}\label{3.4}
Let $(M,g)$ be a semi-Riemannian manifold. If $(M,g)$ has a mixed
3-Sasakian structure
$(\phi_\alpha,\xi_\alpha,\eta_\alpha)_{\alpha=\overline{1,3}}$, then
$(\phi_\alpha)_{\alpha=\overline{1,3}}$ cannot be Killing tensor
fields.
\end{prop}
\begin{demo}
If $(\phi_1,\xi_1,\eta_1,g)$ is a Sasakian structure,
from (\ref{12}) we obtain:
$(\nabla_X\phi_1)X=g(X,X)\xi_1\neq0\nonumber$
for any non-lightlike vector field $X$ orthogonal to $\xi_1$.

For a LP-Sasakian structure $(\phi_2,\xi_2,\eta_2,g)$, from
(\ref{13}) we have:
\begin{equation}
(\nabla_X\phi_2)X=g(\phi_2X,\phi_2X)\xi_2\neq0 \,,
\end{equation}
for any non-lightlike vector field $X$ orthogonal to $\xi_2$.
\end{demo}

\begin{theor}
Let $(M,g)$ be a semi-Riemannian manifold. If $(M,g)$ admits a mixed
3-Sasakian structure, then any conformal Killing vector field on
$(M,g)$ is a Killing vector field.
\end{theor}
\begin{demo}
A vector field $X$ on $M$ is conformal Killing iff
$       L_Xg=f\cdot g $
for $f\in C^\infty(M,\mathbb{R})$ and we have
$(L_Xg)(\xi_\alpha,\xi_\alpha)=fg(\xi_\alpha,\xi_\alpha)=\epsilon_\alpha 
f.$
But, by the Lie operator's properties,
\begin{eqnarray}
&~&(L_Xg)(\xi_\alpha,\xi_\alpha)=Xg(\xi_\alpha,\xi_\alpha)-2g(L_X\xi_\alpha,
\xi_\alpha)
=-2g([X,\xi_\alpha],\xi_\alpha)\nonumber\\
&&=-2g(\nabla_X\xi_\alpha,\xi_\alpha)+2g(\nabla_{\xi_\alpha}X,\xi_\alpha)
=2\epsilon_\alpha g(\phi_\alpha X,\xi_\alpha)-2g(X,\nabla_{\xi_\alpha}
\xi_\alpha)
=0 \,,
\end{eqnarray}
because $\phi_\alpha X \perp \xi_\alpha$ $(\alpha\in\{1,2,3\})$ and
$\nabla_{\xi_\alpha}\xi_\alpha=0$.

Consequently,
$ f=\epsilon_\alpha (L_Xg)(\xi_\alpha,\xi_\alpha)=0 $,
so that $L_Xg=0,$ i.e. $X$ is Killing vector field.
\end{demo}

\end{document}